\documentclass[letterpaper,10pt]{article}
\usepackage{amssymb}
\usepackage[figuresright]{rotating}
\usepackage{hyperref}
\usepackage{wrapfig}
\usepackage{caption}
\usepackage{titling}

\newcommand{\ebartoebarosc}{\mbox{$\bar{\nu_{e}} \rightarrow \bar{\nu_e}$}}
\newcommand{\mutoeosc}{\mbox{$\nu_{\mu} \rightarrow \nu_e$}}
\newcommand{\mutomuosc}{\mbox{$\nu_{\mu} \rightarrow \nu_{\mu}$}}
\newcommand{\degs}{\mbox{$^{\circ}$}}
\newcommand{\mysth}{\mbox{$\mathrm{sin}^2(2\theta_{13})$}}
\newcommand{\sthetaonetwo}{\mbox{$\mathrm{sin}^2(\theta_{12})$}}
\newcommand{\thetatwothree}{\mbox{$\theta_{23}$}}
\newcommand{\sthetatwothree}{\mbox{$\mathrm{sin}^2(2\theta_{23})$}}

\newcommand{\deltacp}{\mbox{$\delta_{CP}$}}

\newcommand{\dmsq}{\mbox{$\Delta m^{2}_{21}$}}
\newcommand{\Dmsq}{\mbox{$\Delta m^{2}_{31}$}}

\title{Precision Neutrino Oscillation Measurements using Simultaneous High-Power, Low-Energy Project-X Beams}
\author{M.Bishai, M.Diwan, S.Kettell, J.Stewart, B.Viren, E.Worcester \\
Brookhaven National Laboratory \and
R.Tschirhart \\ Fermi National Accelerator Laboratory \and
L.Whitehead \\ University of Houston }
\date{}

\begin{document}

\maketitle

\begin{abstract}
The first phase of the long-baseline neutrino experiment, LBNE10, will
use a broadband, high-energy neutrino beam with a 10-kt liquid argon TPC
at 1300 km to study neutrino oscillation. 
In this paper, we describe potential upgrades to LBNE10 that use Project X 
to produce high-intensity, low-energy neutrino beams. 
Simultaneous, high-power 
operation of 8- and 60-GeV beams
with a 200-kt water Cerenkov detector would provide
sensitivity to $\mutoeosc$ oscillations at the second oscillation maximum.
We find that with ten years of data, it would be possible to measure
$\mysth$ with precision comparable to that expected from reactor antineutrino
disappearance and to measure the value of the CP phase, $\deltacp$, 
with an uncertainty of 
$\pm(5-10)\degs$.
This document is submitted for inclusion in Snowmass 2013.
\end{abstract}

\vspace{0.5in}

Recent measurements of non-zero $\mysth$\cite{dayabay_a,reno,double-chooz,
dayabay_b} enable the search for CP violation in the neutrino sector 
and, ultimately,
precision measurement of the CP phase, $\deltacp$, using $\mutoeosc$
oscillations.
The first phase of the long-baseline neutrino experiment, LBNE10, 
will use 708 kW of 120-GeV protons from Fermilab's Main
Injector (MI) to produce a muon-neutrino or antineutrino beam aimed 
at a 10-kt liquid argon time projection chamber (LAr TPC) 
at a distance of 1300 km. 
The spectrum of neutrino energies detected at the far 
site in LBNE10 is aligned with the first oscillation maximum, peaking in the 
range E$_{\nu}$ = (2-4)~GeV.
As described in its conceptual design report\cite{LBNE-CDR}, LBNE10, 
in combination with other
neutrino data, is expected to determine the neutrino
mass hierarchy and provide an
initial measurement of the CP phase in the three-generation framework.

\begin{figure}[htbp]
\centering
\includegraphics[width=0.8\textwidth]{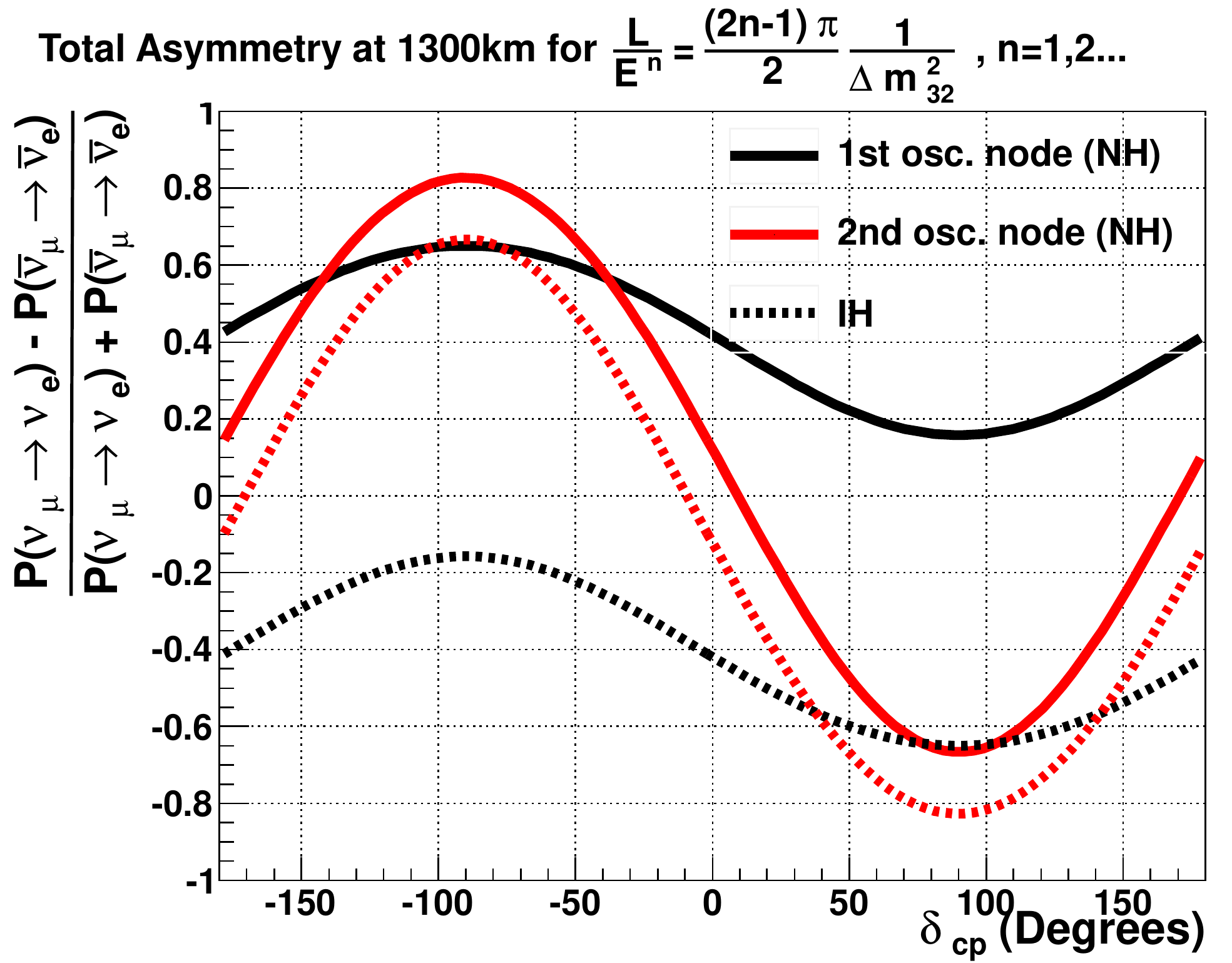}
\caption{Total neutrino-antineutrino asymmetry in the probability of
$\mutoeosc$ appearance at 1300 km, at the first (black line) and second
(red line) oscillation maxima, for normal (solid line) and inverted 
(dashed line) hierarchy, as a function
of the true value of $\deltacp$.}
\label{fig:oscprob}
\end{figure}

Figure~\ref{fig:oscprob} shows the total neutrino-antineutrino asymmetry
in the probability of 
$\mutoeosc$ appearance as a function of $\deltacp$, at the first and second
oscillation peaks, for normal and inverted hierarchy, at a distance
of 1300 km. This asymmetry includes contributions from both CP and matter 
effects.
It is clear from Fig.~\ref{fig:oscprob}
that the matter effect is large in the first oscillation maximum, but 
in the second oscillation maximum, where E$_{\nu}$ = (0.2-1.5)~GeV, the
CP asymmetry is large with very little matter asymmetry. For this reason,
measurement of $\mutoeosc$ appearance at the second oscillation maximum 
provides excellent sensitivity to CP violation, independent of 
the mass hierarchy.

Project X\cite{ProjectX} will make it possible to produce
high-intensity, low-energy neutrino beams. 
In this paper, we summarize and update \cite{NOPE}, 
which argues that simultaneous, high-power operation of 8- and 60-GeV beams
with a 200-kt water Cerenkov detector at a long baseline provides
sensitivity to $\mutoeosc$ oscillations at the second oscillation maximum,
allowing precise measurements of neutrino oscillation 
parameters independent of the mass hierarchy.

The kinematics of neutrino beam production dictates that the only way
to produce significant yield of neutrinos at low energies is with high 
proton-beam
power at low energies. With Project X, 
beam power from the MI can be maintained
at or above 2 MW for proton energies of 60~GeV to 120~GeV. 
Upgrading to an 8-GeV pulsed LINAC would provide up to ~4 MW of 8-GeV beam
power, only 270 kW of which is required by the MI to
produce the 2-MW, 60-GeV beam.
In this scenario, the Fermilab accelerator complex could simultaneously
produce 2 MW of 60-GeV protons and 3 MW of 8-GeV protons. 
The resulting neutrino beams would have significant flux
with E$_{\nu} < 1.5$~GeV, which would allow measurement of 
$\mutoeosc$ oscillation at the second oscillation maximum at 1300 km.

Here, we consider a 200-kt water Cerenkov detector, with reconstruction 
performance similar to Super-Kamiokande (SK)\cite{superk_atmospheric}, 
located at
Sanford Underground Research Facility (SURF)\cite{surf}, 
as a potential future upgrade to LBNE. 
The efficiency 
of a water Cerenkov detector for quasielastic neutrino scattering, 
which dominates at low neutrino energy, is $\sim$80\%.  
For this study, we have used GLoBES\cite{globes1,globes2} to estimate
the experimental sensitivities using the oscillation parameters,
constraints (where applicable), and experimental assumptions shown in
Table~\ref{tab:globesin}.
The central values of and constraints on the oscillation parameters
are taken from a global fit to experimental
neutrino data\cite{Fogli2012}.

\begin{table}
\centering
\caption{Summary of the oscillation paramaters, constraints, and 
experimental assumptions that were used in the  GLoBES calculations
presented in this paper.}
\begin{tabular}[c]{|l|c|c|} \hline
Parameter & Central Value & Uncertainty ($1\sigma$) \\ \hline
$\sthetaonetwo$ & 0.31 & 5\% \\
$\thetatwothree$ & 38.3$\degs$ & 8\% \\
$\mysth$ & 0.094 & 5\% \\
$\dmsq$ & $+7.5 \times 10^{-5} \mathrm{eV}^2$ & 3\% \\
$\Dmsq$ (NH) & $+2.5 \times 10^{-3} \mathrm{eV}^2$ & 3\% \\ 
Matter density & 2.8 g/cm$^3$ & 2\% \\
Signal normalization & n/a & 1\% \\
Background normalization & n/a & 5\% \\ \hline
\end{tabular}
\label{tab:globesin}
\end{table}

\begin{figure}[htbp]
\centering
\includegraphics[width=0.49\textwidth]{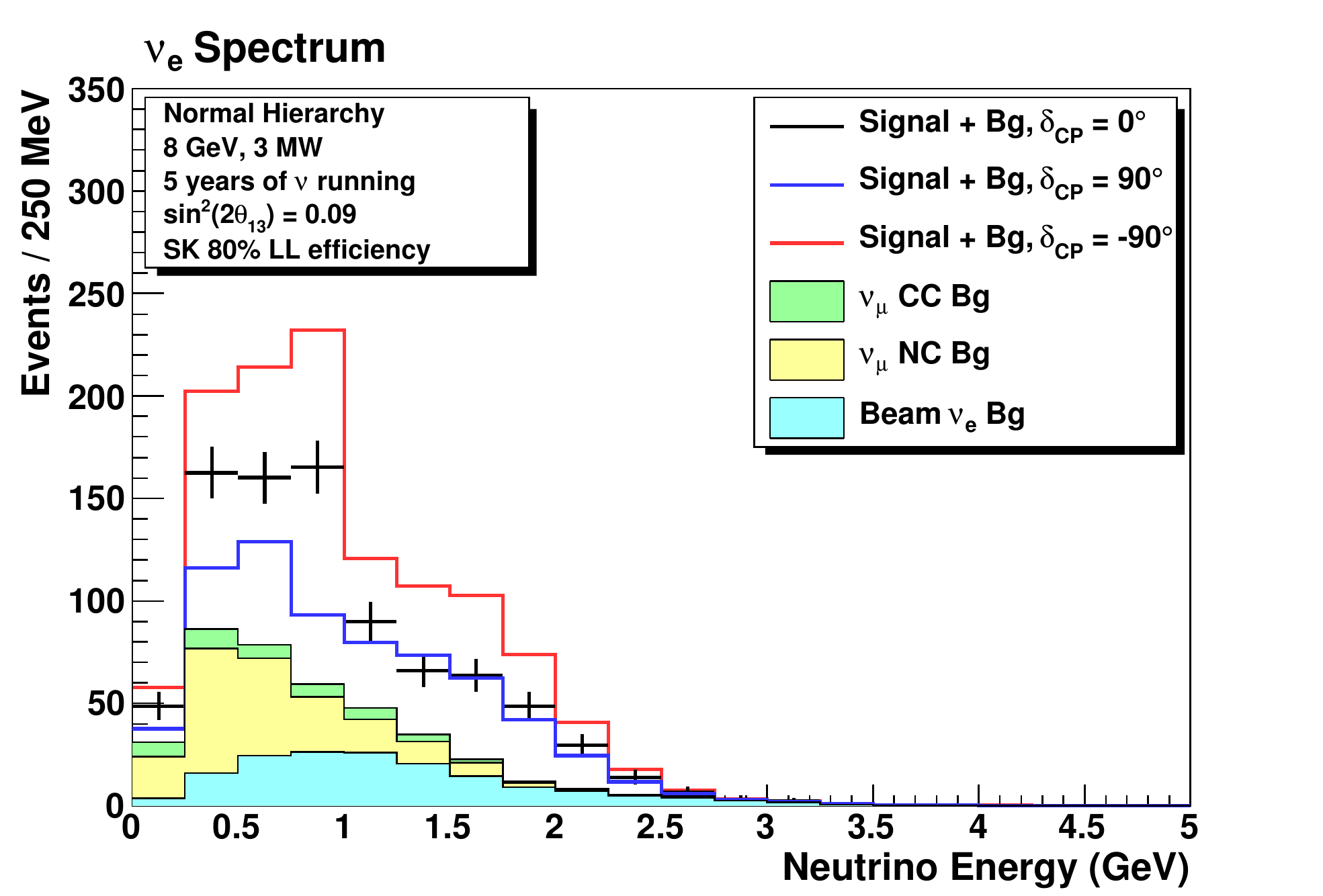}
\includegraphics[width=0.49\textwidth]{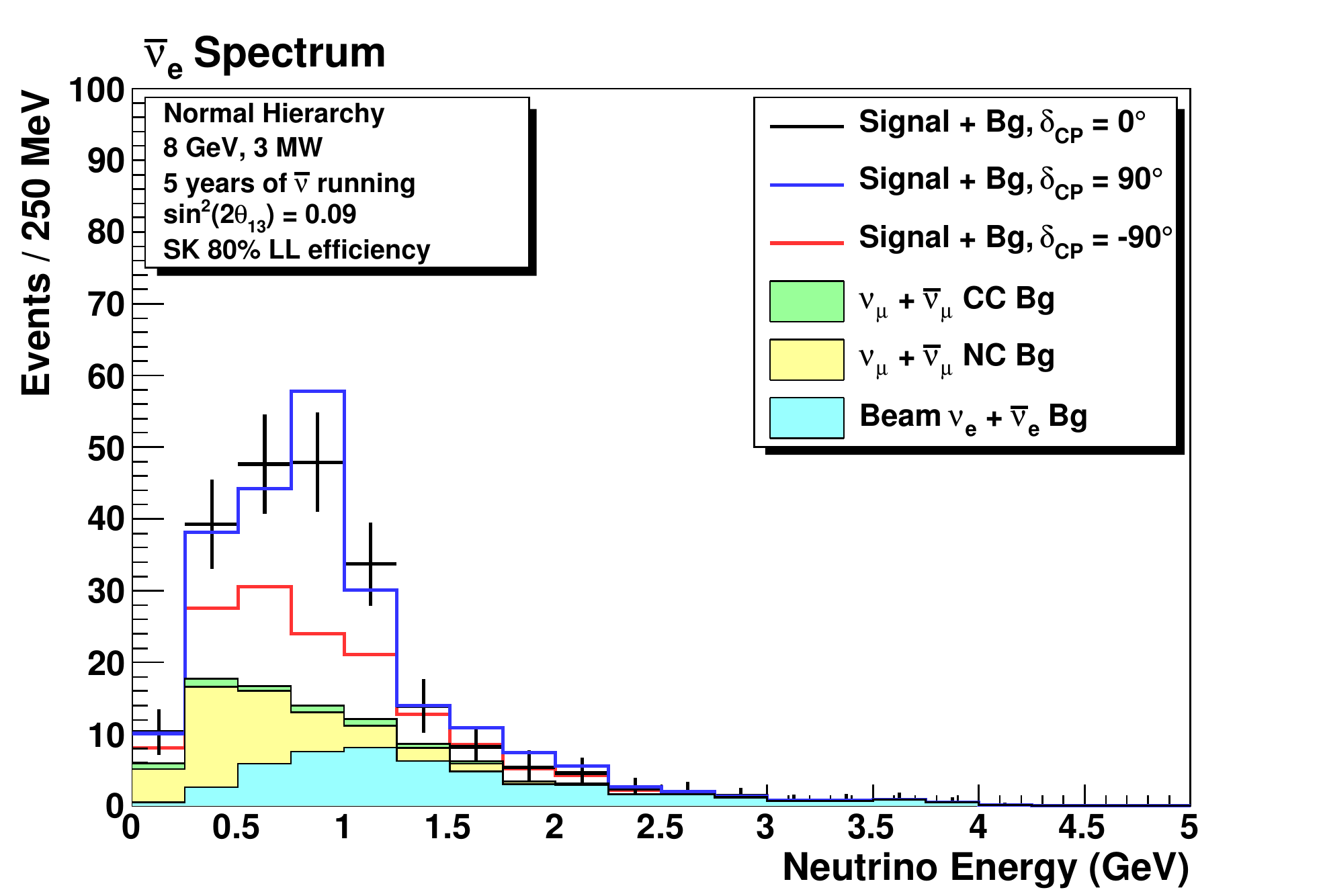}
\includegraphics[width=0.49\textwidth]{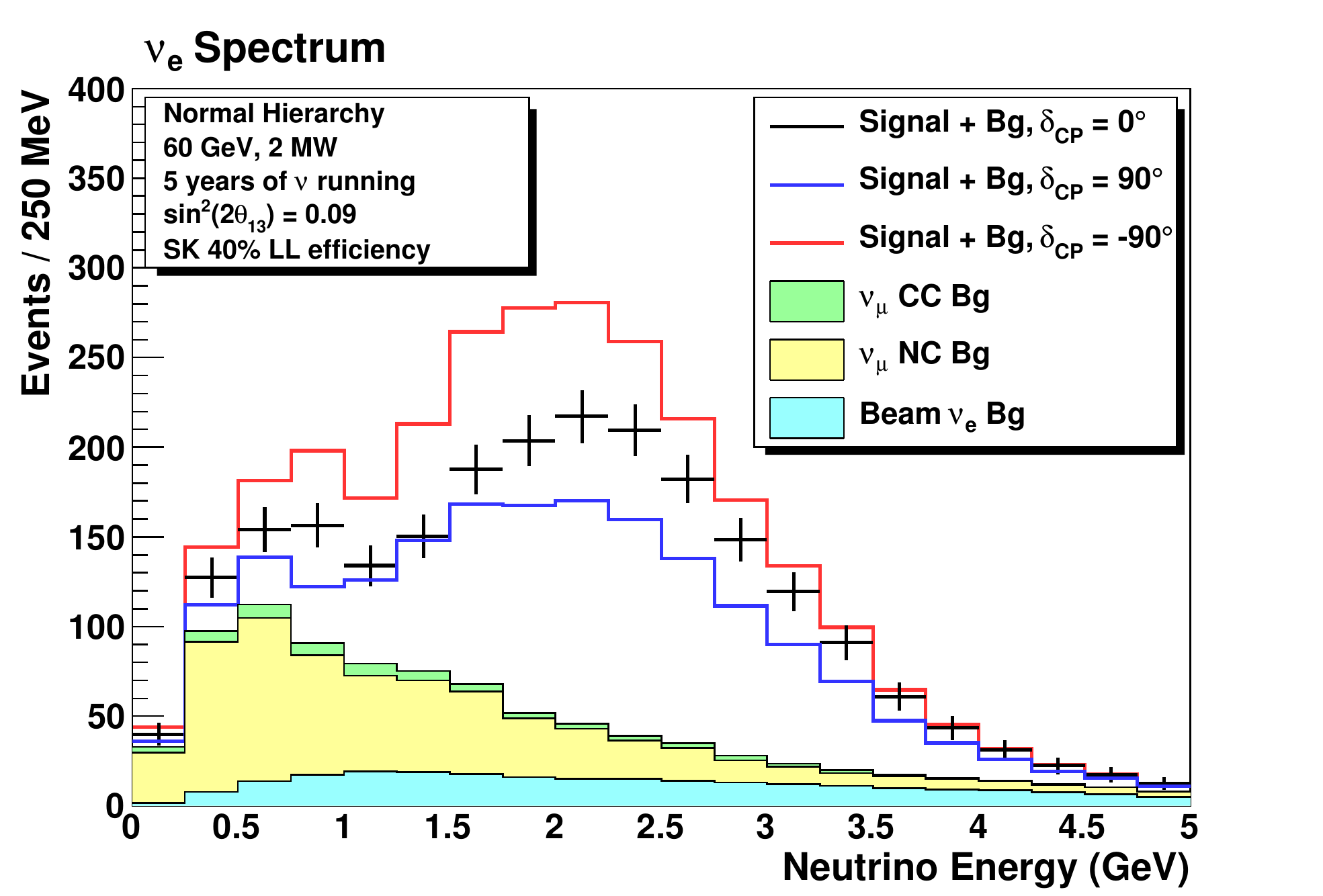}
\includegraphics[width=0.49\textwidth]{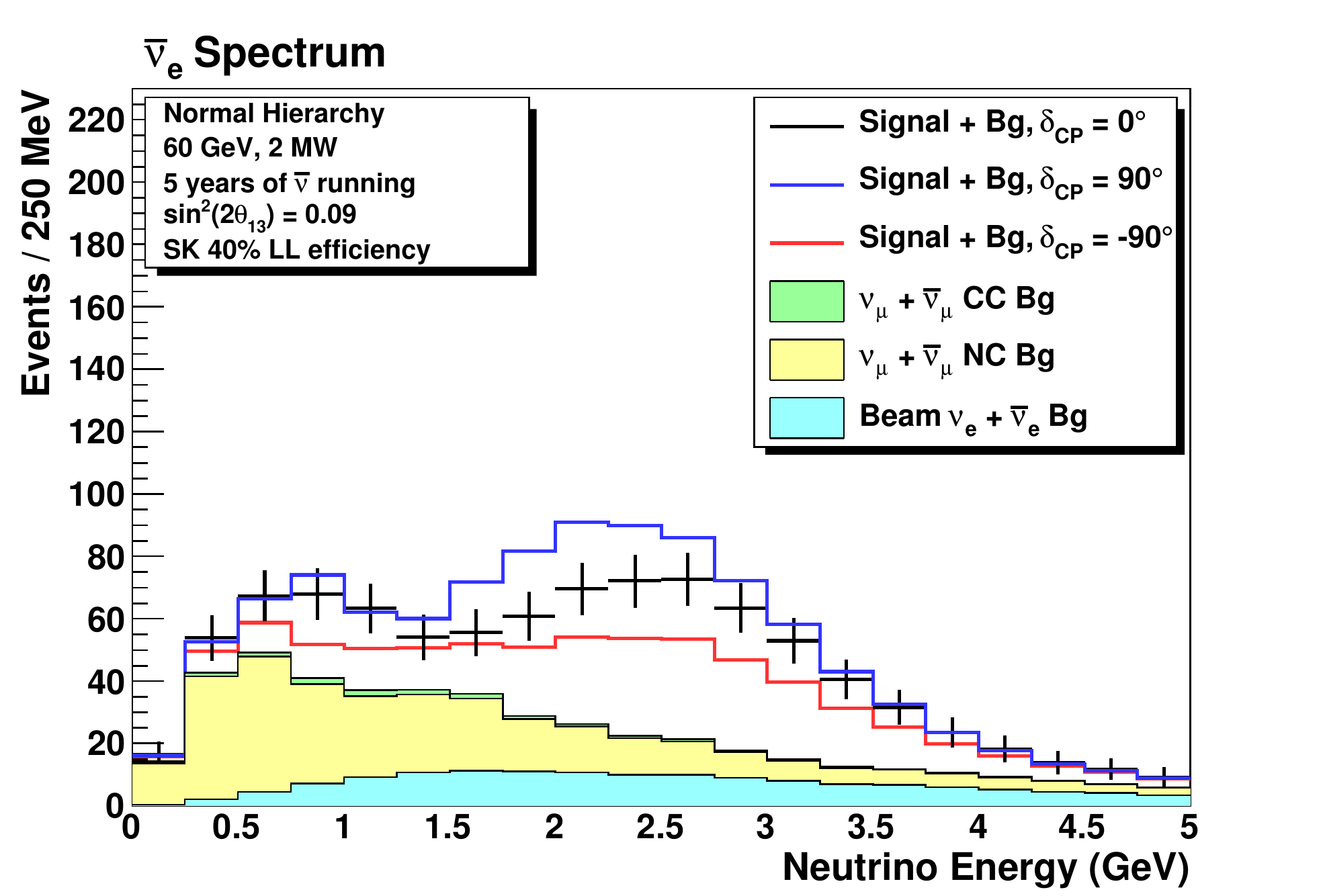}
\caption{Expected signal and background distributions for neutrinos (left)
and antineutrinos (right) for the 8-GeV (top) and 60-GeV (bottom) beams.
The exposure is 5 years each for neutrinos and antineutrinos, with 3 MW of
beam power for the 8-GeV beam and 2 MW of beam power for the 60-GeV beam.
Normal hierarchy is assumed. The signal distributions for three possible
values of $\deltacp$ are shown.
Efficiencies for the 8-GeV and 60-GeV data are calculated
assuming SK 80\% and 40\% log-likelihood selection criteria, respectively.}
\label{fig:spectra}
\end{figure}

SK uses a log-likelihood (LL) variable to distinguish between 
$\nu_e$ charged-current signal and background.   
Log-likelihood variable cuts can be chosen such that 
there is 40\% signal efficiency with little background or 80\% 
signal efficiency with higher background levels.
We have applied the SK 80\% LL efficiencies for the 8-GeV beam, where 
background is expected to be low, and 40\% LL efficiencies for the 60-GeV
beam in which the background level is higher. Selection criteria could, of
course, be optimized further, but we do not find large changes in the 
experimental sensitivity from tightening or loosening these cuts.
Figure \ref{fig:spectra} shows the spectra for five years of
neutrino and antineutrino running with the 8-GeV and 60-GeV beams, 
assuming normal hierarchy.
The low-energy background level 
is significantly reduced relative to higher energy beams.

With ten years of running in simultaneous 8- and 60-GeV mode with a
200-kt water Cerenkov detector,
in combination with data from LBNE10,
it would be possible to measure both $\mysth$ and $\deltacp$ 
very precisely. For the results shown here, we assume that the
8-GeV data is all in neutrino mode and the 60-GeV data is equally
divided between neutrinos and antineutrinos. We do not include
the contribution from additional 
data that could be taken with the LBNE10 LAr TPC during
the ten years of low-energy running.

\begin{figure}[htbp]
\centering
\includegraphics[width=0.9\textwidth]{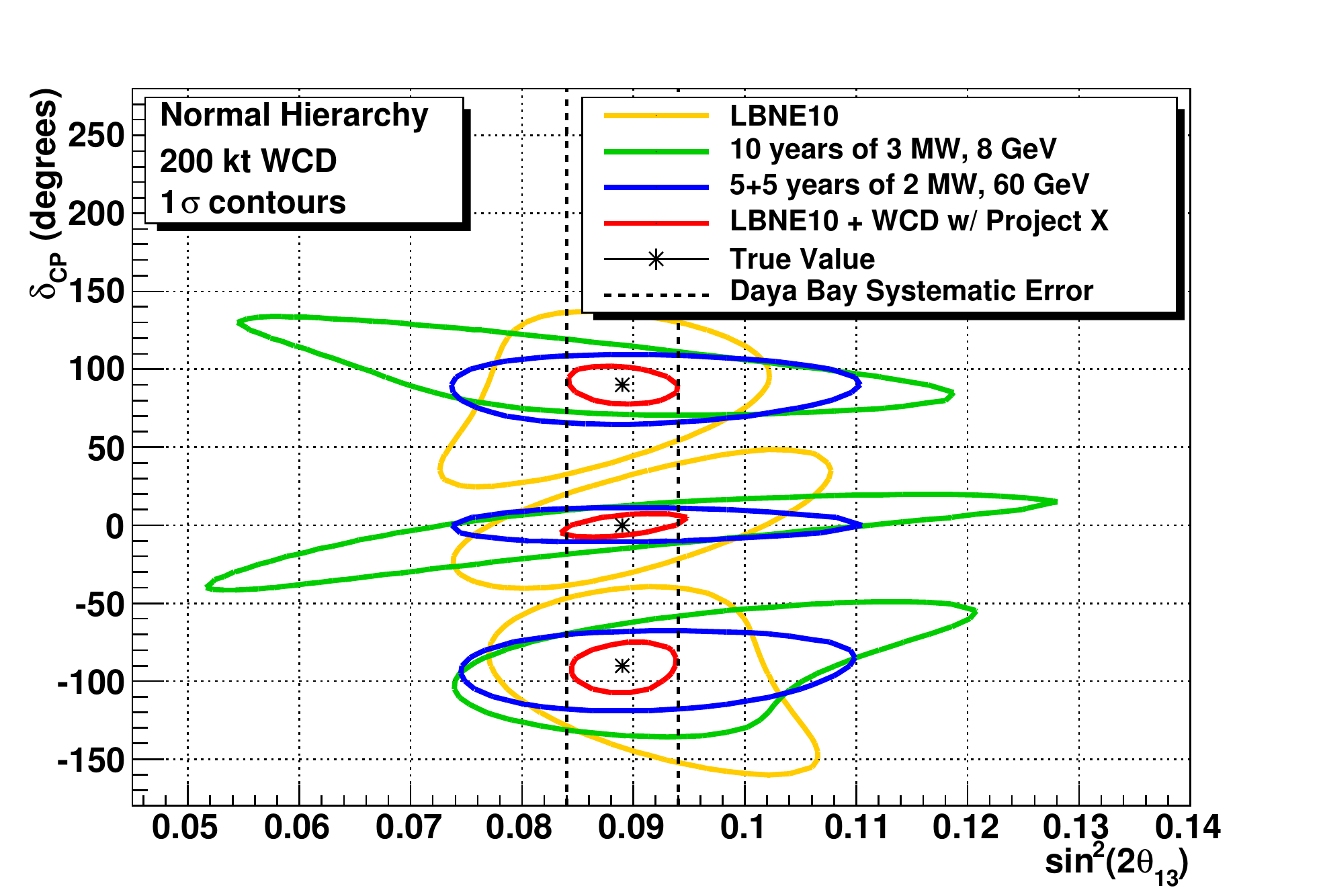}
\caption{One-sigma contours from a two-dimensional fit for $\mysth$
and $\deltacp$ for LBNE10(gold), 60-GeV data(blue), 8-GeV data(green),
and the combination of the three(red). Fit results for three possible
true values of $\deltacp$ are shown. The Daya Bay systematic error on
$\mysth$ (dashed lines) is shown for comparison.} 
\label{fig:bubbles}
\end{figure}

The $1\sigma$ contours from a two-dimensional
fit to $\mysth$ and $\deltacp$ for LBNE10, 60-GeV data, 8-GeV data,
and the combination of the three are shown in Fig.~\ref{fig:bubbles}.
The precision
on $\mysth$, coming primarily from the 120-GeV
and 60-Gev data, is competitive with 
the precision expected from reactor neutrino experiments. 
The measurement of $\mysth$ is complementary
to that from reactor neutrino experiments because it is measured in
$\mutoeosc$ oscillations rather than $\ebartoebarosc$ disappearance.
The 8-GeV and 60-GeV data are highly sensitive to 
$\deltacp$. The combination of 120-GeV, 60-GeV, and 8-GeV data
provides a precise measurement of $\deltacp$ with no
external constraint on $\mysth$. 
Additionally, the 8-GeV, 60-GeV, and 120-GeV data
place independent constraints on neutrino
oscillation parameters; new physics could be detected as inconsistent
measurements of neutrino oscillation parameters in these three data sets.

\begin{figure}[htbp]
\centering
\includegraphics[width=0.9\textwidth]{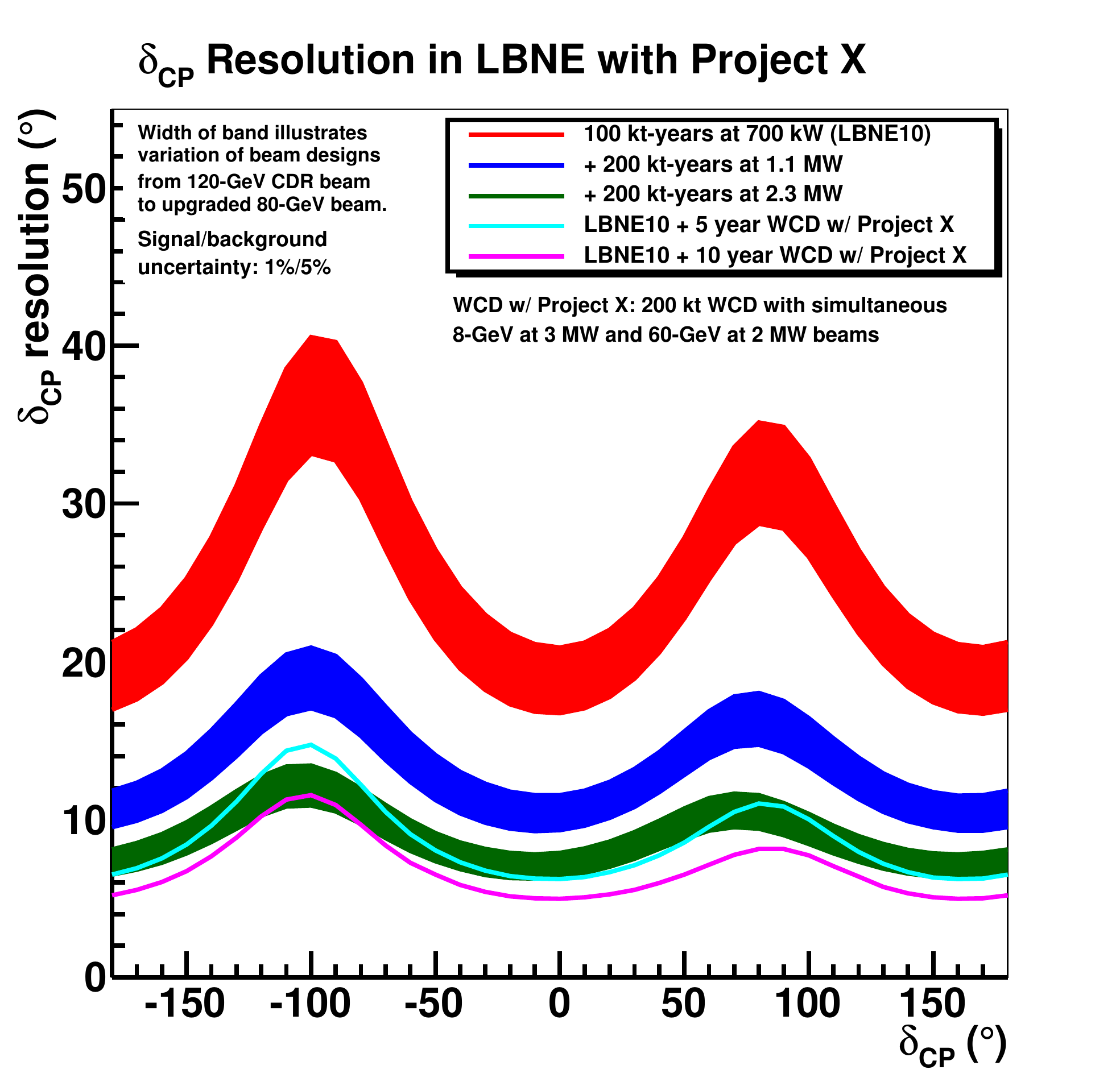}
\caption{Resolution on the measurement of $\deltacp$ as a function
of the true value of $\deltacp$ for various LBNE configurations.
LBNE10 (red) refers to 100 kt-years of exposure to a 80-120 GeV, 700 kW
beam for a LAr TPC. The blue and green curves show the resolution that
can be achieved by a LAr TPC with higher exposures; total exposure for 
the blue curve is 290 kt-MW-years
and for the green curve is 750 kt-MW-years. The cyan and pink curves show
the resolution for LBNE10 in combination with five and ten years, respectively,
of simultaneous, low-energy beams and a 200-kt WCD, as described in this
paper.} 
\label{fig:resolutions}
\end{figure}

Figure \ref{fig:resolutions} compares the one-dimensional
resolution on the measurement
of $\deltacp$ for simultaneous low-energy beams to LBNE LAr TPC detectors
with a range of exposure to beams with energy 80-120 GeV,
varying from 70 kt-MW-years to 750 kt-MW-years.
Here we apply a 5\% constraint on the value of $\mysth$ for all data sets.
This constraint is consistent
with the expected final uncertainty on the 
value of $\mysth$ from reactor
antineutrino disappearance experiments.
This constraint is very important for the
resolution of $\deltacp$ at low exposures; for the simultaneous low-energy
beams, such a constraint is less significant because the internal resolution
on $\mysth$ is comparable to the external constraint.
Ten years of data with a 200-kt WCD and simultaneous low-energy beams, in
combination with LBNE10, provides the best
$\deltacp$ resolution for almost all true values of $\deltacp$, and is the
only configuration considered that can reach $5\degs$ resolution for any
value of $\deltacp$.

\begin{figure}[htbp]
\centering
\includegraphics[width=0.9\textwidth]{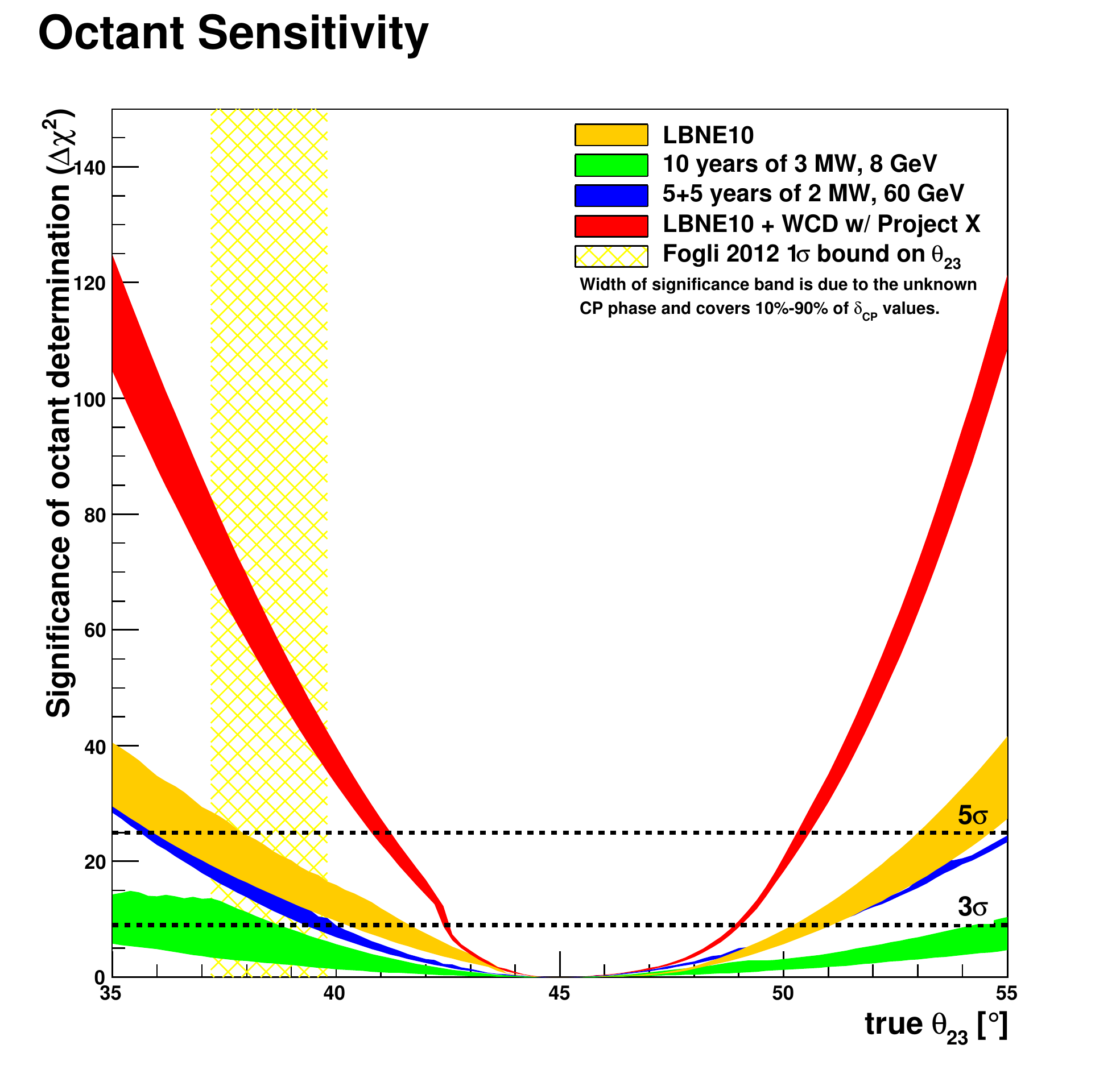}
\caption{Significance of the octant determination ($\Delta \chi^2$ for 
a fit to the ``wrong'' octant) 
as a function of the true value of $\thetatwothree$
 for LBNE10(gold), 60-GeV data(blue), 8-GeV data(green),
and the combination of the three(red). The width of each curve is due to the
unknown CP phase and covers 10\% to 90\% of possible true $\deltacp$ values.
The 1$\sigma$ bound on $\thetatwothree$ from a global fit to neutrino
experimental data\cite{Fogli2012} is shown in yellow for reference.} 
\label{fig:octant}
\end{figure}

Adding the 8- and 60-GeV data to LBNE10 also significantly improves
sensitivity to resolution of the $\theta_{23}$ octant degeneracy.
Figure \ref{fig:octant} shows the significance of the octant determination
for LBNE10, 60-GeV data, 8-GeV data, and the combination of the three, as
a function of the true value of $\thetatwothree$. In this fit, $\mutomuosc$
disappearance contributes to the precision on $\sthetatwothree$ while the
$\mutoeosc$ appearance data provides information on the 
$\thetatwothree$ octant. 
Again, a 5\% external
constraint on the true value of $\mysth$ has been applied; this
is necessary for the LBNE10 analysis, but becomes less important with the
addition of the low-energy data.
A 5$\sigma$ determination
of the octant of $\thetatwothree$ will be possible for at least 90\% of
true values of $\deltacp$ for $41\degs < \thetatwothree_{true} < 51\degs$.
Most of the sensitivity comes from the 120-GeV and 60-GeV data because
the second oscillation maximum does not provide any special sensitivity
to the octant determination.

In summary, the cleanest, most dramatic sensitivity to the CP phase
comes from measurement of $\mutoeosc$ oscillation at the second oscillation
maximum, at long baseline, with a
high-mass far detector. Project X, with an 8-GeV pulsed LINAC, could produce
simultaneous low-energy, high-intensity beams which probe this
low-energy region, making precision measurements
of neutrino oscillation parameters possible.

\bibliographystyle{hieeetr}
\bibliography{pxloe_bib}

\end{document}